\documentclass[useAMS,usenatbib,usegraphicx]{mn2e}

\usepackage{hyperref} \usepackage{color} \usepackage{subfigure}
\usepackage{verbatim} \usepackage{afterpage}
\pdfoutput=1

\title[Nova shell around the cataclysmic variable V1315
Aquilae]{Discovery of an old nova shell surrounding the cataclysmic
  variable V1315 Aql} \author[D. I. Sahman et
al]{D. I. Sahman,$^{1}$\thanks{E-mail: d.sahman@sheffield.ac.uk;
    vik.dhillon@sheffield.ac.uk} V. S. Dhillon,$^{1,2}$\footnotemark[1]
  S. P. Littlefair,$^{1}$ and G. Hallinan $^{3}$\\ $^{1}$Department of
  Physics and Astronomy, University of Sheffield, Sheffield S3 7RH,
  UK\\ $^{2}$Instituto de Astrofisica de Canarias, E-38205 La Laguna,
  Tenerife, Spain\\ ${^3}$California Institute of Technology, 1200 East
  California Boulevard, Pasadena, California 91125, USA\\}

\begin{document}

\date{Accepted 2018 April 12. Received 2018 March 19}

\pagerange{\pageref{firstpage}--\pageref{lastpage}} \pubyear{2013}

\maketitle

\label{firstpage}

\begin{abstract}

  Following our tentative discovery of a faint shell around V1315 Aql
  reported in \citet{sahman15}, we undertook deep H$\alpha$ imaging
  and intermediate-resolution spectroscopy of the shell. We find that
  the shell has its geometric centre located on V1315 Aql. The mass,
  spectral features and density of the shell are consistent with other
  nova shells, rather than planetary nebulae or supernova
  remnants. The radial velocity of the shell is consistent with the
  systemic velocity of V1315 Aql. We believe this evidence strongly
  suggests that the shell originates from an earlier nova event. This
  is the first nova shell discovered around a novalike, and supports
  the theory of nova-induced cycles in mass transfer rates
  (hibernation theory) first proposed by \citet{shara86}.

\end{abstract}

\begin{keywords} stars:  -- novae -- cataclysmic variables.
\end{keywords}

\section{Introduction}

Cataclysmic variables (CVs) are close binary systems in which a white
dwarf (WD) primary accretes material from a late-type secondary star,
via Roche-lobe overflow (see \citealt{warner95a} for a
review). Non-magnetic CVs are classified into 3 main sub-types -- the
novae, the dwarf novae and the nova-likes. The {\em novae} are defined
as systems in which only a single nova eruption has been
observed. Novae eruptions have typical amplitudes of 10 magnitudes and
are believed to be due to the thermonuclear runaway of hydrogen-rich
material accreted onto the surface of the white dwarf. The {\em dwarf
  novae} (DNe) are defined as systems which undergo quasi-regular (on
timescales of weeks--months) outbursts of much smaller amplitude
(typically 6 magnitudes). Dwarf novae outbursts are believed to be due
to instabilities in the accretion disc causing it to collapse onto the
white dwarf. The {\em nova-like} variables (NLs) are the non-eruptive
CVs, i.e. objects which have never been observed to show novae or
dwarf novae outbursts. The absence of dwarf novae outbursts in NLs is
believed to be due to their high mass-transfer rates, producing
ionised accretion discs in which the disc-instability mechanism that
causes outbursts is suppressed \citep{osaki74}; the mass transfer
rates in NLs are $\dot{M} \sim 10^{-9}$ M$_{\odot}$ yr$^{-1}$ whereas
DNe have rates of $\dot{M} \sim 10^{-11}$ M$_{\odot}$ yr$^{-1}$
\citep{warner95a}.

Our understanding of CV evolution has made great strides in recent
years (e.g. \citealt{knigge10}, \citealt{knigge11}).  However, one of
the main unsolved problems in CV evolution is: how can the different
types of CV co-exist at the same orbital period?  Theory predicts that
all CVs evolve from longer to shorter orbital periods on timescales of
gigayears, and as they do so the mass-transfer rate also declines. At
periods longer than approximately 5 hours, all CVs should have high
mass-transfer rates and appear as nova-likes, whereas below this
period the lower mass-transfer rate allows the disc-instability
mechanism to operate and all CVs should appear as dwarf novae
\citep{knigge11}. This theoretical expectation, however, is in stark
contrast to observations, which show that nova-likes are far more
common than dwarf novae in the 3--4 hr period range
\citep{rodriguez07}.

Two possible explanations for the coexistence of nova-likes and dwarf
novae at the same orbital periods have been proposed, both of which
invoke cycles in $\dot{M}$ on timescales shorter than the gigayear
evolutionary timescale of CVs. The first explanation is that the
$\dot{M}$ cycles are caused by irradiation from the accreting WD,
which bloats the secondary and hence increases $\dot{M}$
(e.g. \citealt{buning04}). \citet{knigge11} found that irradiation
would cause bloating of ${<}3\%$ above the period gap, leading to
modest fluctuations in $\dot{M}$ with timescales of the order of
$10^6-10^9$ yr, insufficient to explain the full range in $\dot{M}$
that is observed. The second explanation for variable $\dot{M}$ is a
nova-induced cycle. Some fraction of the energy released in the nova
event will heat up the WD, leading to irradiation and subsequent
bloating of the secondary. Following the nova event, the system would
have a high $\dot{M}$ and appear as a NL. As the WD cools, $\dot{M}$
reduces and the system changes to a DN, or even possibly $\dot{M}$
ceases altogether and the system goes into hibernation. Hence CVs are
expected to cycle between nova, NL and DN states, on timescales of
$10^4-10^5$ yrs (see \citealt{shara86}).

The cyclical evolution of CVs through nova, NL and DN phases recently
received observational support from the discovery that BK Lyn appears
to have evolved through all three phases since its likely nova
outburst in the year AD 101 \citep{patterson13}. A second piece of
evidence has come from the discovery of nova shells around the dwarf
novae Z Cam and AT Cnc (\citealt{shara07}, \citealt{shara13}),
verifying that they must have passed through an earlier nova
phase. \citet{shara17} also found a nova shell from Nova Sco 1437 and
were able to associate it with a nearby dwarf nova using its proper
motion. A more obvious place than DNe to find nova shells is actually
around NLs, as the nova-induced cycle theory suggests that the high
$\dot{M}$ in NLs is due to a recent nova outburst. Finding shells
around NLs would lend further support to the existence of nova-induced
cycles and hence why systems with different $\dot{M}$ are found at the
same orbital period.

In our earlier paper (\citealt{sahman15}; hereafter S15), we presented
the initial results of our search for nova shells around CVs. We
reported the tentative discovery of a possible shell around the
nova-like V1315 Aql (orbital period 3.35 hr). We subsequently obtained
intermediate-resolution spectroscopy of this shell, in an effort to
determine its physical characteristics and to ascertain if it is
associated with the nova-like. The results of these spectroscopic
observations, along with a more in-depth analysis of the H$\alpha$
images of the V1315 Aql shell shown in S15, are presented in this
paper.

\section{Observations and Data Reduction}

\subsection{Observations}

\subsubsection{INT images}

We used the Wide Field
Camera\footnote{http://www.ing.iac.es/astronomy/instruments/wfc/} at
the prime focus of the 2.5m Isaac Newton Telescope on La Palma to
image V1315 Aql on the night of 2014~August~2. This setup gave a
platescale of 0.33$\arcsec$/pixel and a field view of approx.
$34\arcmin \times 34\arcmin$. H$\alpha$ is generally the strongest
feature in the spectra of nova shells, with a velocity width of up to
2000 km\,s$^{-1}$ (e.g.  \citealt{duerbeck87}). In order to maximise
the detection of light from the shell and minimise the contribution of
sky, we therefore used a narrow-band (95\AA\ FWHM = 4300 km\,s$^{-1}$)
interference filter centred on the rest wavelength of H$\alpha$ (ING
filter number
197\footnote{http://catserver.ing.iac.es/filter/list.php?instrument=WFC}). We
took eight 900s H$\alpha$ exposures, with four of the images dithered
by $\pm 20\arcsec$ in both RA and Dec. The observing conditions were
good throughout the run: the sky was always photometric, there was no
evidence of dust and the seeing was 1.5$\arcsec$.

\subsubsection{Keck DEIMOS spectra}

We used the DEIMOS \citep{faber03} multi-slit spectrograph on the 10m Keck II
telescope on Hawaii, on the night of 2015~June~13. We obtained 39
spectra of 300s duration each, using the 1200G grating centred on
6000\AA\,\, and the GG455 order-blocking filter. This gave a
wavelength coverage of 4550--7500\AA, with a FWHM
resolution of 1.6\AA. The seeing was 0.7$\arcsec$, and there was some
thin cloud present.

The slit mask design requires that the slits cannot overlap in the
spatial direction, so we placed seven slits around the edges of the
roughly circular shell. We also placed a slit on V1315 Aql itself and
chose four nearby stars for flux calibration. We identified two areas
of blank sky for sky subtraction. The positions of each slit on the
sky are shown in Figure \ref{fig:mask1}, and full details of the
position, orientation and wavelength coverage of each slit are given
in Table \ref{tab:journal}.

\begin{table*}
\caption[]{V1315 Aql DEIMOS slit positions, sizes and spectral range
  coverage. The RA and Dec positions are for the centres of the slits.}
\begin{center}
\begin{tabular}{lllccccc}
\hline
 \multicolumn{1}{l}{Slit name} & \multicolumn{1}{l}{RA} & \multicolumn{1}{l}{Dec}
  & \multicolumn{1}{c}{Slit} & \multicolumn{1}{c}{Slit} &
 \multicolumn{1}{c}{Slit} & \multicolumn{2}{c}{Wavelength range} \\
& \multicolumn{1}{l}{(degs)} & \multicolumn{1}{l}{(degs)} & \multicolumn{1}{c}{Length} &
\multicolumn{1}{c}{Position} & \multicolumn{1}{c}{Width} &
   \multicolumn{1}{c}{Start(\AA)} & \multicolumn{1}{c}{End (\AA)} \\ 
& & & \multicolumn{1}{c}{(arcsecs)} & angle &
 \multicolumn{1}{c}{(arcsecs)}  & & \\
& & & & (degs) & & & \\ \hline Blank\,\,Sky 1 & 288.5230602 & 12.2217710 &
58.872 & 154.4 & 0.7 & 4780 & 7432 \\ Blank Sky 2 & 288.4896981 &
12.3217610 & 49.423 & 154.4 & 0.7 & 4867 & 7514 \\ Shell 1 &
288.4995975 & 12.3003892 & 32.904 & 154.4 & 1.0 & 4868 & 7521 \\ Shell
2 & 288.5116182 & 12.2922447 & 46.933 & 150.0 & 1.0 & 4915 & 7576
\\ Shell 3 & 288.4560525 & 12.3369014 & 74.181 & 170.0 & 1.0 & 4686 &
7342 \\ Shell 4 & 288.4507472 & 12.3566674 & 65.419 
& 170.0 & 1.0 & 4699 & 7343 \\ Shell 5 & 288.5171327 & 12.2647206 &
59.595 & 154.4 & 1.0 & 4877 & 7529 \\ Shell 6 & 288.4757487 &
12.2611586 & 43.389 & 150.0 & 1.0 & 4614 & 7272 \\ Shell 7 &
288.4367229 & 12.3103574 & 43.386 & 130.0 & 1.0 & 4487 & 7170 \\ V1315
Aql & 288.4769928 & 12.3013719 & 42.173 & 154.4 & 1.0 & 4735 & 7382
\\ Star 1 & 288.5492156 & 12.2172214 & 49.155 & 154.4 & 1.0 & 5058 &
7565 \\ Star 2 & 288.5403078 & 12.2461305 & 45.629 & 154.4 & 1.0 &
5047 & 7592 \\ Star 3 & 288.5135945 & 12.2472213 & 42.009 & 154.4 &
1.0 & 4898 & 7455 \\ Star 4 & 288.5604882 & 12.2053443 & 61.154 &
154.4 & 1.0 & 5038 & 7586 \\
\hline
\end{tabular}
\end{center}
\label{tab:journal}
\end{table*}

\begin{figure*}
  \centering \includegraphics[width=140mm,angle=0]{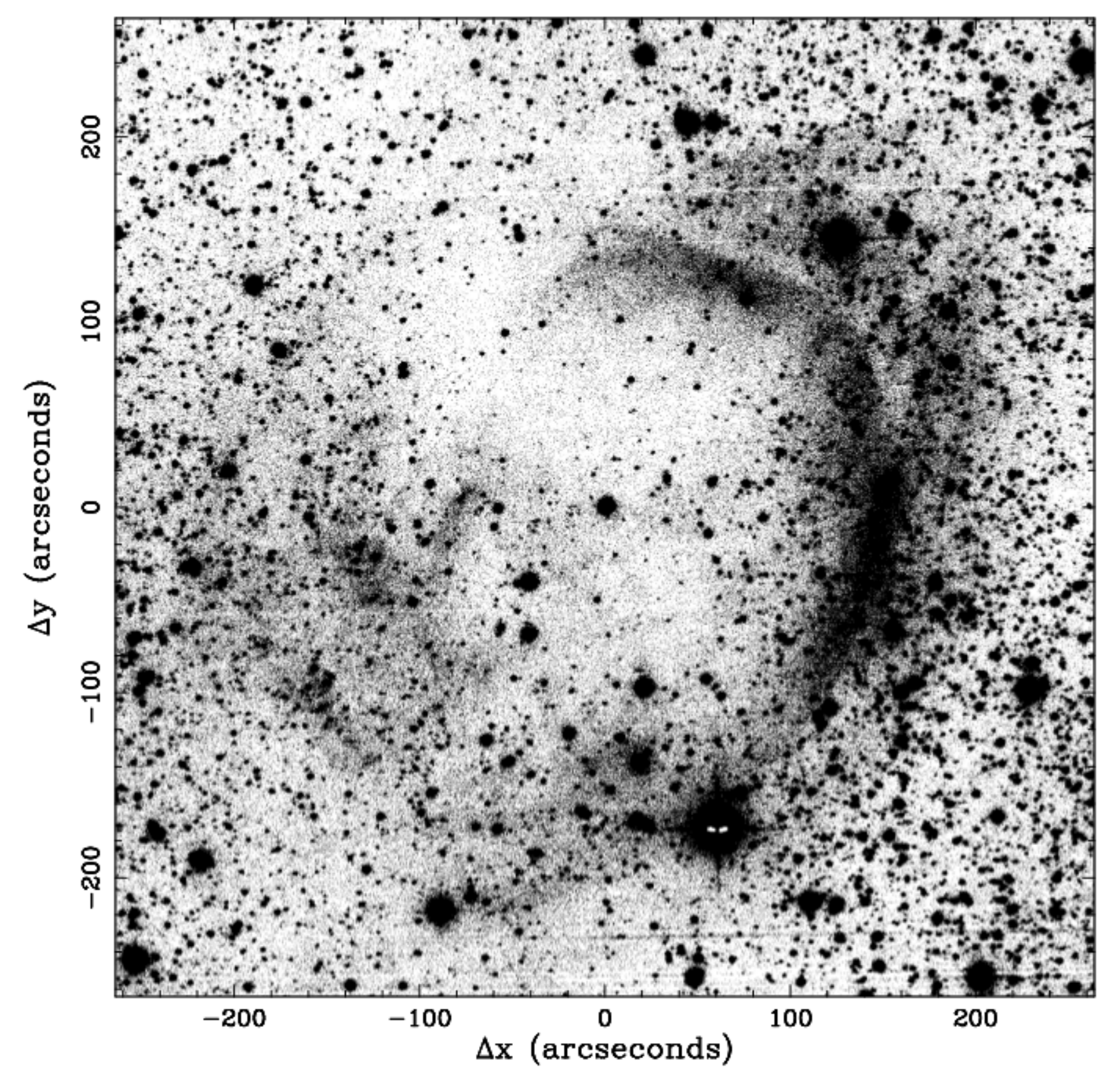}
\caption{INT WFC H$\alpha$ image of the nova shell around V1315
  Aql. The binary is located at the centre of the image. North is up
  and East is left.}
\label{fig:zoom}
\end{figure*}

\begin{figure*}
  \centering \includegraphics[width=140mm,angle=0]{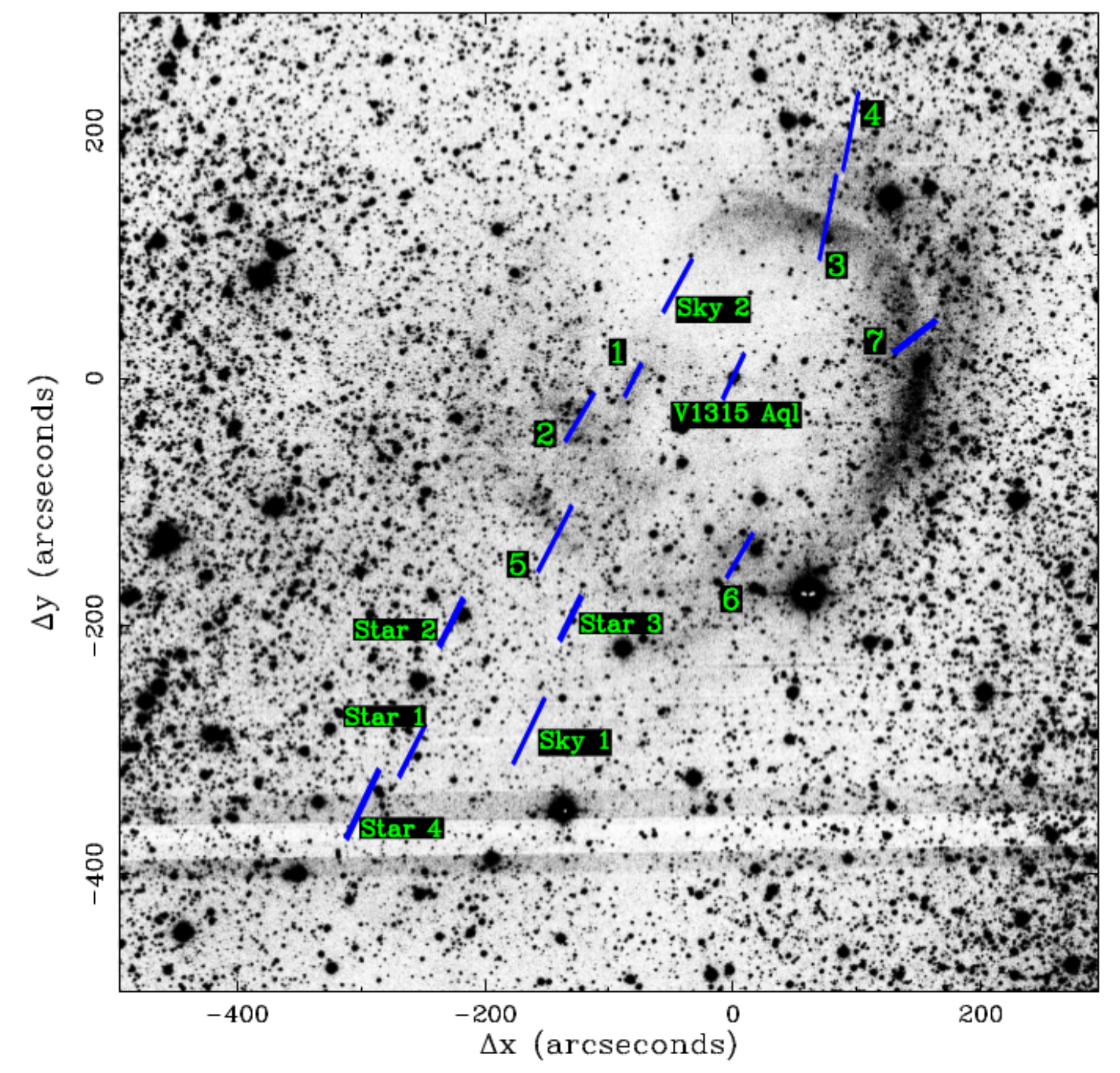}
  \caption{INT WFC H$\alpha$ image of the nova shell around V1315 Aql
    with the Keck DEIMOS slit positions and sizes overlaid. V1315 Aql
    is situated at \textit{x}=0, \textit{y}=0. The seven shell slits
    are numbered, and the two blank sky slits are also shown. The four
    flux calibration stars are marked Star 1--4. North is up and East
    is left. The horizontal band across the image at \textit{y}$\sim
    -380$ is the gap between two of the CCDs in the WFC mosaic.}
\label{fig:mask1}
\end{figure*}

\subsubsection{0.5m Telescope -- La Palma}

Our Keck spectra included four stars for flux calibration but
unfortunately they did not appear in any photometric catalogues.  In
order to allow us to perform flux calibration, we therefore obtained
additional images of the four stars together with two catalogue stars
(TYC 1049-408-1 and IPHAS J1911411.93+121357.7) using the 0.5m robotic
telescope \textit{pt5m} on La Palma \citep{hardy15}.  The observations
were taken on 2016 October 7, when we took four images in each of the
\textit{B, V, R, I} filters with an exposure time of 1 minute each,
and on 2016 Nov 18, when we took four 40 sec \textit{R}-band images
and four 360 sec \textit{B}-band images.

\subsection{Data reduction}

\subsubsection{INT images}

The INT images were debiased using the median level of the overscan
strip and flat-fielded using normalised twilight sky flats. All image
processing was carried out using {\sc
  theli}\footnote{http://www.ing.iac.es/astronomy/instruments/wfc/WFC-THELI-reduction.html}.
Figure \ref{fig:zoom} shows the final stacked image of the shell.

\subsubsection{Keck DEIMOS spectra}

We used {\sc iraf} to reduce the DEIMOS spectra. The spectra were bias
corrected using the overscan strip on the chips, and were flat-fielded
using quartz lamp flats. We had difficulty in performing the
background sky subtraction because the two blank sky slits we had
chosen both contain small residual $H\alpha$ emission lines, possibly
from the nova shell.

We then tried using the sky portion of our four flux calibration
stars, but we found that the spectra of the three closest to the shell
(Stars 1--3) also contained low levels of residual $H\alpha$
emission. (See Figure \ref{fig:haspec}). The best results were
obtained with sky from Star 4, which is furthest from the shell and
showed negligible $H\alpha$ emission -- this was used for all
subsequent background sky subtraction.

\subsubsection{pt5m images}

The images were bias and flat field corrected using standard {\sc
  iraf} procedures. This allowed us to derive magnitudes for the four
flux calibration stars, as shown in Table \ref{tab:mags}. We also
found 2MASS infrared magnitudes \citep{skrutskie06} for Star 2. Hence,
the star with the most complete set of magnitudes was Star 2. We input
these values into the Virtual Observatory SED Analyser (VOSA -- see
\citealt{bayo08}) to determine the spectral type of Star 2, obtaining
M4V ($\pm 2$). We then used VOSA to generate a template spectrum of an
M4V star, which we used to flux calibrate the Keck spectra in {\sc
  iraf}.

\begin{table}
  \caption[]{Magnitudes of the four flux-calibration stars observed
    with Keck DEIMOS. The errors on the \textit{B, V, R, I} magnitudes
    are $\pm0.3$ magnitudes. See Table \ref{tab:journal} for the
    positions of the stars on the sky.}
\begin{center}
\begin{tabular}{lrrrrr}
  \hline
  \multicolumn{1}{l}{Band} & \multicolumn{1}{r}{Star 1} &
\multicolumn{1}{r}{Star 2} & \multicolumn{1}{r}{Star 3} &
\multicolumn{1}{r}{Star 4} & \multicolumn{1}{r}{} \\ \hline \textit{B} & 18.8 &
16.5 & -- & -- \\ \textit{V}  & 17.6 & 15.3 & 19.5 & 18.3 \\ \textit{R} & 16.9 &
14.5 & 18.4 & 16.5 \\ \textit{I} & 15.1 & 13.7 & 17.2 & 14.6 \\ 2MASS \textit{J}
& -- & 12.508 & -- & -- \\ 2MASS \textit{H} & -- & 11.798 & -- & -- \\ 2MASS \textit{K}
& -- & 11.646 & -- & -- \\
\hline
\end{tabular}
\end{center}
\label{tab:mags}
\end{table}

\subsubsection{Review of satellite imagery}

We searched the GALEX UV satellite footprint using the GalexView
interface (\citealt{bianchi14}), but no observations were taken of the
field around V1315 Aql. We also examined the WISE 22$\mu$m data
(\citealt{wright10}), and there was no emission in the vicinity of
V1315 Aql.

\section{Results}
\label{sec:res}

\subsection{INT image}

The H$\alpha$ image of the shell surrounding V1315 Aql is shown in
Figure \ref{fig:zoom}. The images clearly show one, possibly two
roughly spherical shells centred on V1315 Aql. The lobe towards the
West has the most prominent emission. There was no evidence of
nebulosity on wider scales than shown in Figure \ref{fig:mask1}.

There is a possibility that the shell is unrelated to V1315 Aql, and
it may just be a line-of-sight alignment of a foreground or background
cloud of gas in the Milky Way. To determine that the shell does indeed
originate from V1315 Aql, we need to determine if it has the same
systemic velocity as the binary, and that its composition is
comparable to other nova shells, and to rule out other types of
nebulosity, e.g. planetary nebulae, supernovae remnants.

\subsection{Geometry of the shell}

\begin{figure*}
\centering \includegraphics[width=140mm,angle=0]{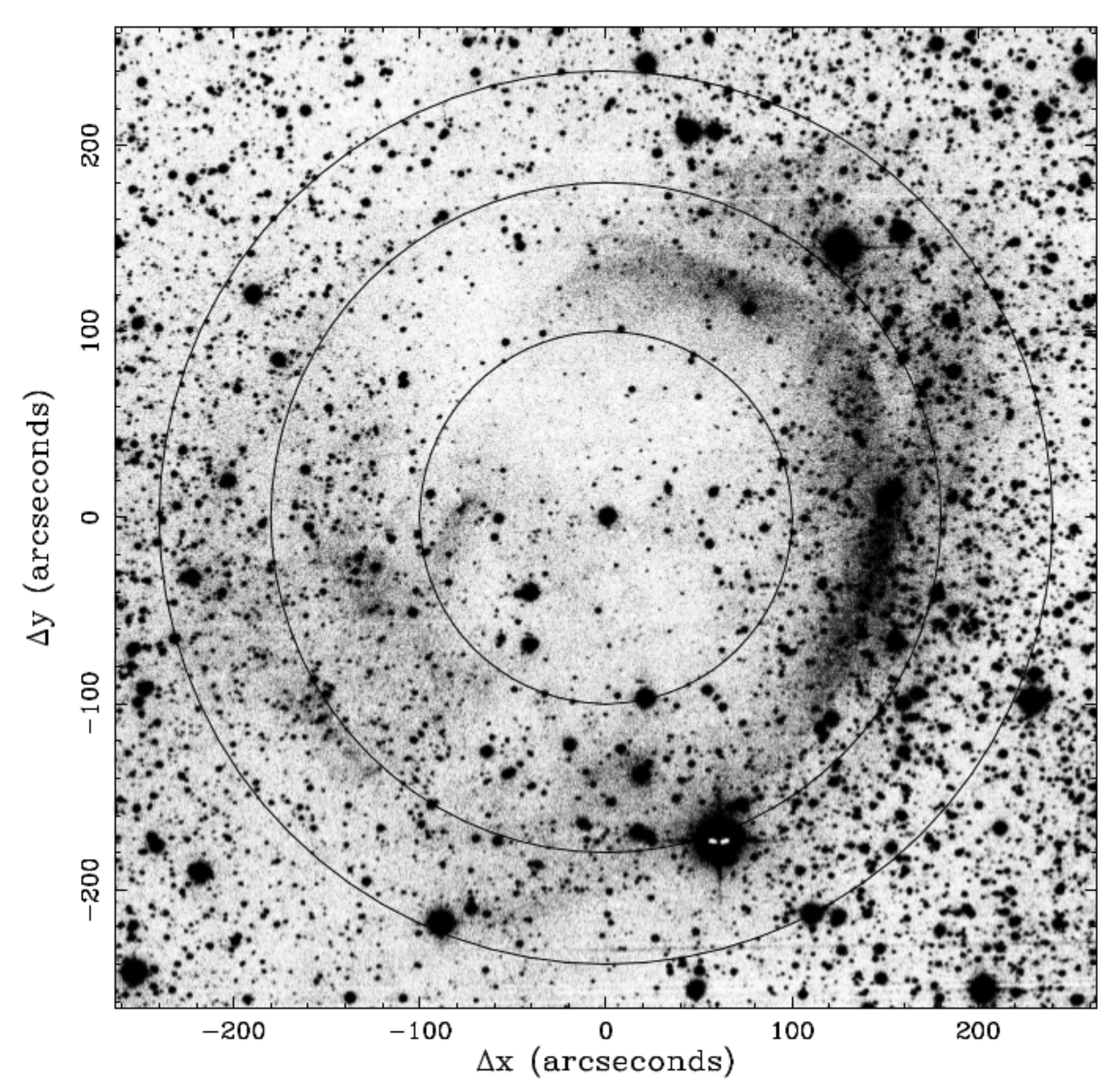}
\caption{INT WFC H$\alpha$ image of the shell with overlaid circles
  centred on V1315 Aql of radii 100\arcsec, 180\arcsec and
  240\arcsec. North is up and East is left.}
\label{fig:geocirc}
\end{figure*}

In Figure \ref{fig:geocirc} we show the image of the shell with
circles centred on V1315 Aql overlaid. The radii of the circles are
100\arcsec, 180$\arcsec$ and 240\arcsec. The inner annulus between
100$\arcsec$ and 180$\arcsec$ contains the most prominent areas of
emission (from the North around to the West), and appears to be
centred on V1315 Aql. The outer annulus also contains a fainter arc of
emission to the North-West, and some fainter areas of emission to the
South-East, which also appear to be centred on V1315 Aql.

\subsection{Keck DEIMOS spectra}

\begin{figure}
\centering \includegraphics[width=85mm,angle=0]{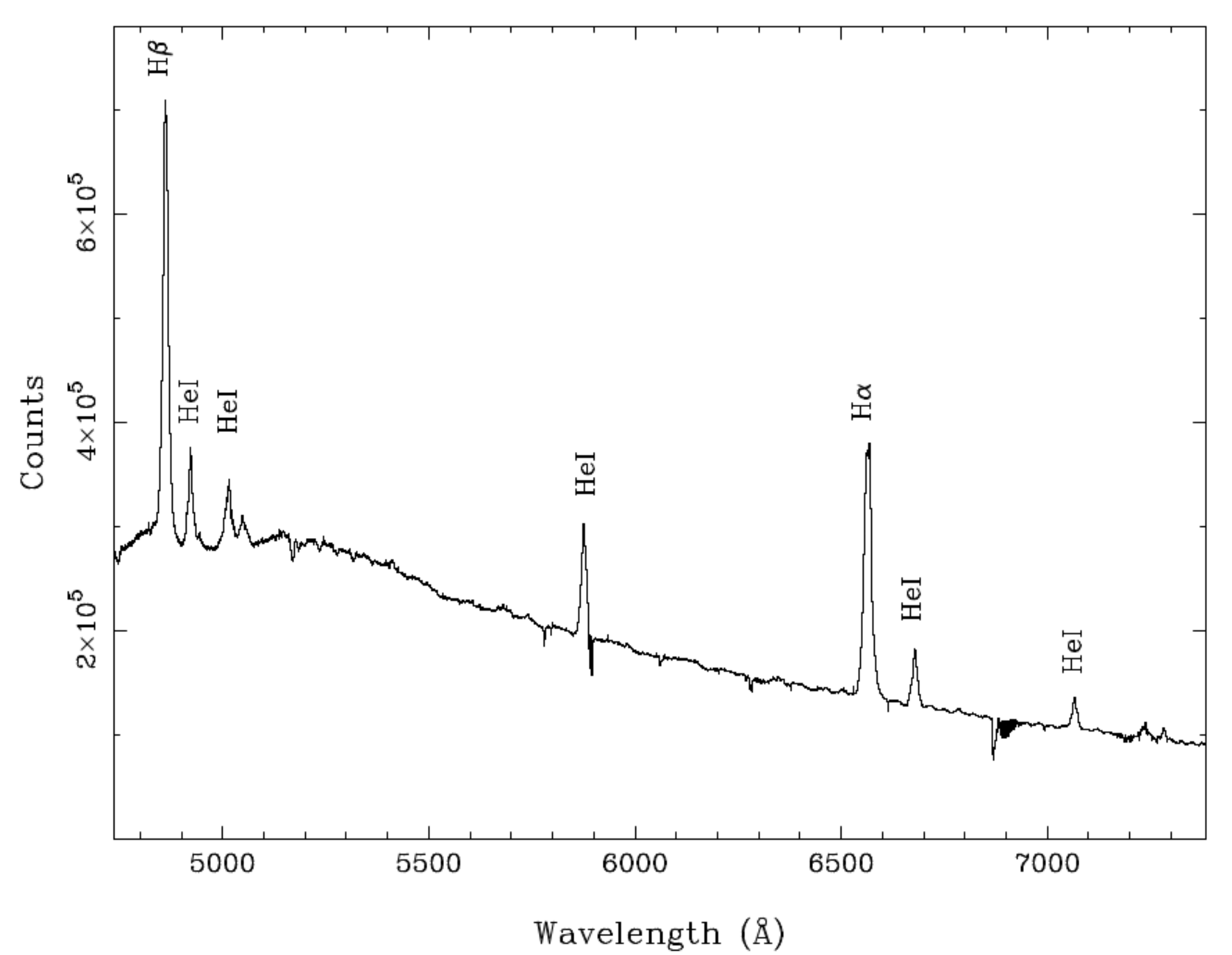}
\caption{Keck DEIMOS spectrum of V1315 Aql. Note that we did not flux
  calibrate this spectrum because the flux calibration stars do not
  cover its whole wavelength range.}
\label{fig:vtot}
\end{figure}

In Figure \ref{fig:vtot} we show the spectrum of V1315 Aql. The
spectrum shows strong, broad (FWHM of H$\alpha$ is 900 km s$^{-1}$)
Balmer and He{\sc{I}} emission lines from the accretion disk. The
spectrum is very similar to that shown in \citet{dhillon95b}.

\subsubsection{Emission lines}
\label{emline}

\begin{figure*}
\centering \includegraphics[width=180mm,angle=0]{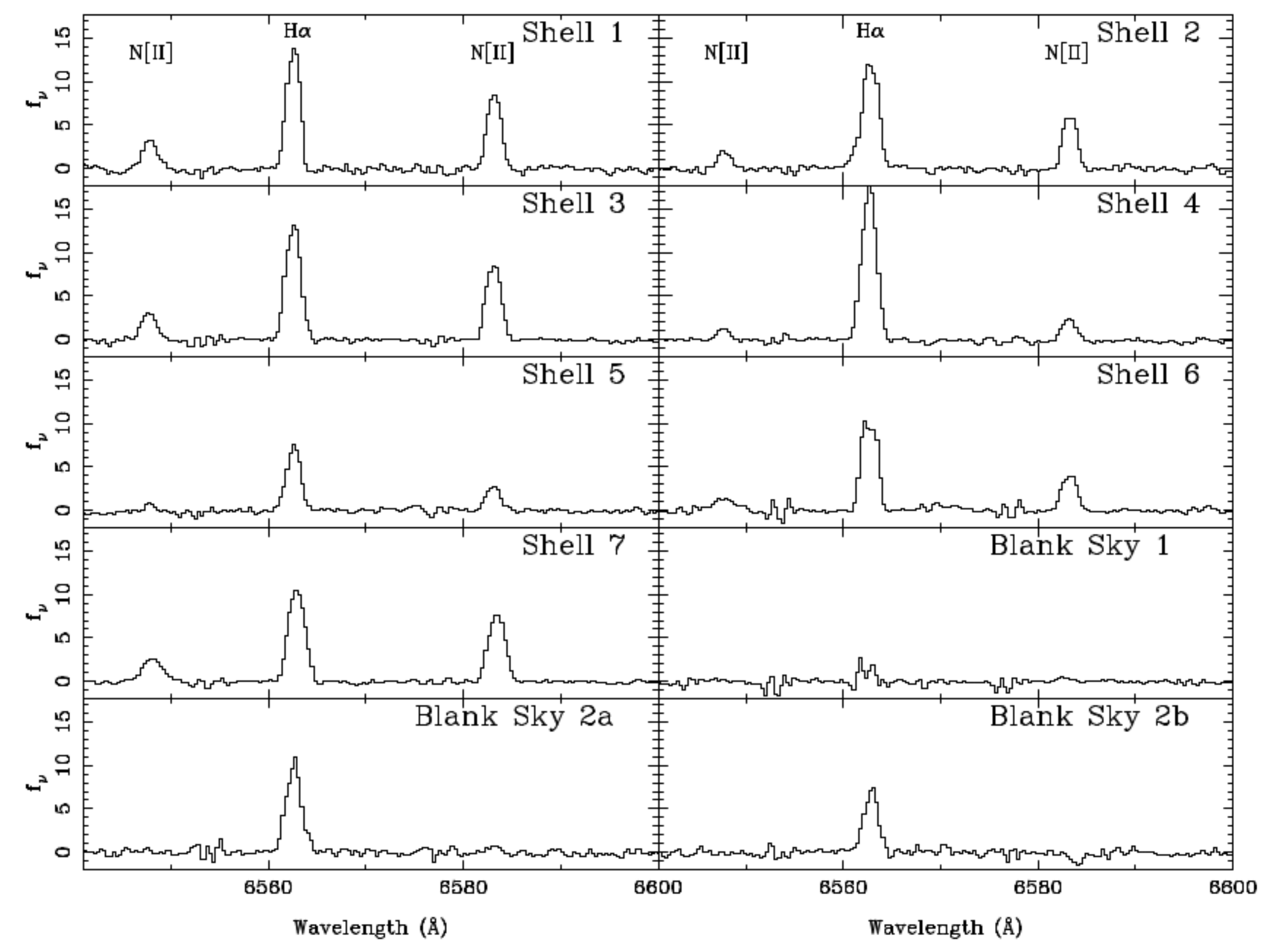}
\caption{Spectra of the seven shell slits (see Fig. \ref{fig:mask1}),
  and the blank sky slits from 6540--6600\AA. The flux is in units of
  $\mu$Jy arcsec$^{-2}$. The slit spectra all show the presence of
  H$\alpha$ and N[II] 6548\AA\,\,and N[II] 6583\AA. H$\alpha$ is also
  present in the blank sky slits. The Blank Sky 2 slit fell across two
  CCDs on the detector and so we show each spectrum separately, as 2a
  and 2b.}
\label{fig:haspec}
\end{figure*}

The spectra of the seven shell slits and the blank sky slits in the
range 6540--6600\AA\, are shown in Fig. \ref{fig:haspec}. Note that
the blank sky 2 slit spanned two CCDs in the spectrograph and each
part is shown separately. The shell spectra all show single-peaked
emission lines of H$\alpha$ and a pair of N[II] lines at 6548 and 
6583\AA. These lines are characteristic of old nova shells
\citep{downes01}.

\begin{figure*}
\centering \includegraphics[width=150mm,angle=0]{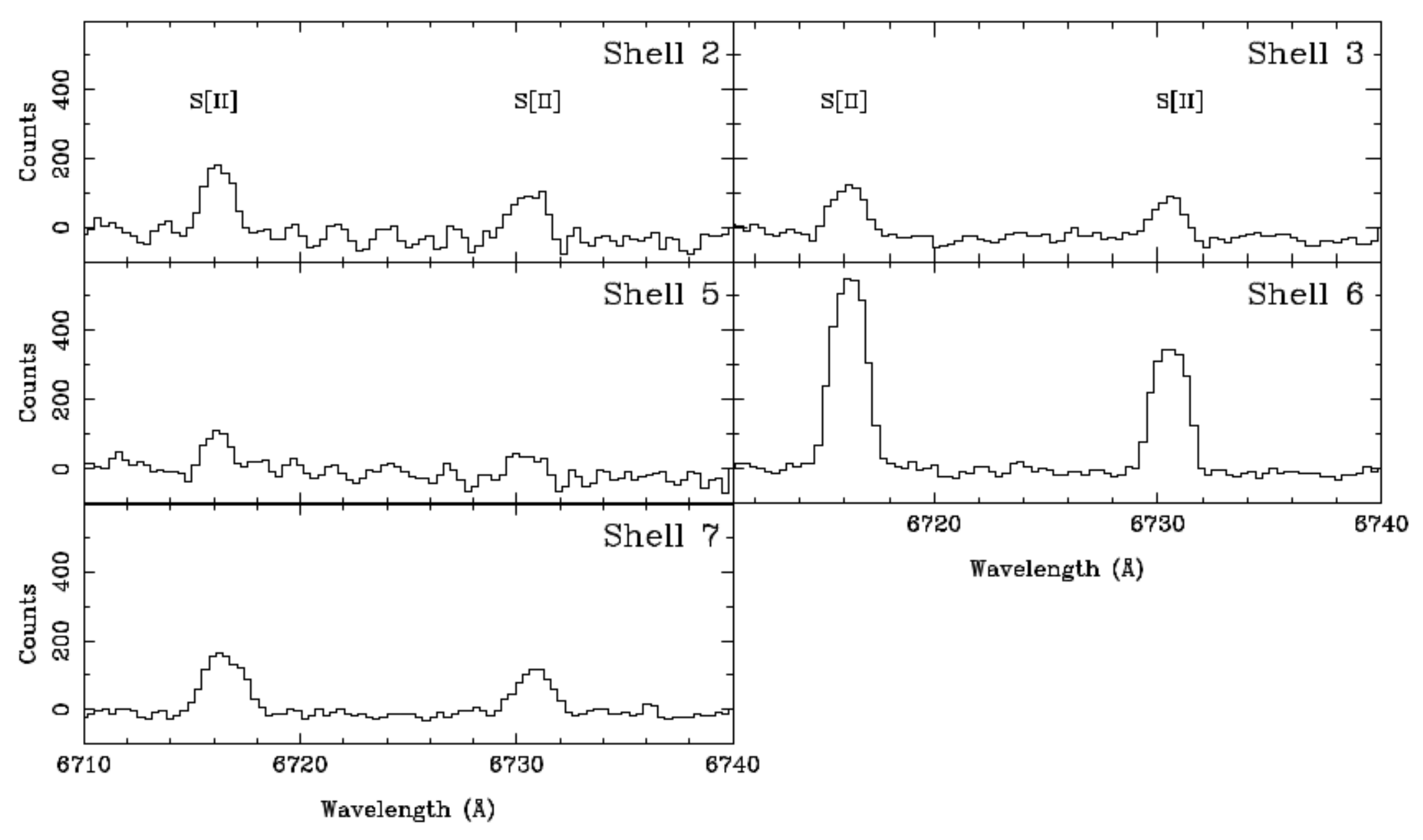}
\caption{Spectra of the five shell slits from 6710--6740\AA\, showing
  the S[II] emission lines at 6716 and 6731\AA.}
\label{fig:siispec}
\end{figure*}

We also detected H$\beta$ in those shell spectra that covered
4861\AA. Unfortunately, none of the spectra of the four
flux-calibration stars covered this wavelength, and hence we were
unable to flux calibrate the H$\beta$ lines. We also found the S[II]
6716 and 6731\AA\, lines in shell slits 2, 3, 5, 6 and 7, as shown in
Figure \ref{fig:siispec}. The average ratio of the two S[II] lines is
1:1.4.

We searched for the emission lines N[II] 5755\AA\,, O[I] 6300, 6300,
6364\AA\, and O[III] 4363, 4959, 5007\AA, often seen in nova shell
spectra, but none were detected. There are faint lines at 5679, 5740
and 5742\AA, presumably from NI and NII, in shell slits 3 and 4, but
these are not present in any other slits.

\begin{figure}
\centering \includegraphics[width=85mm,angle=0]{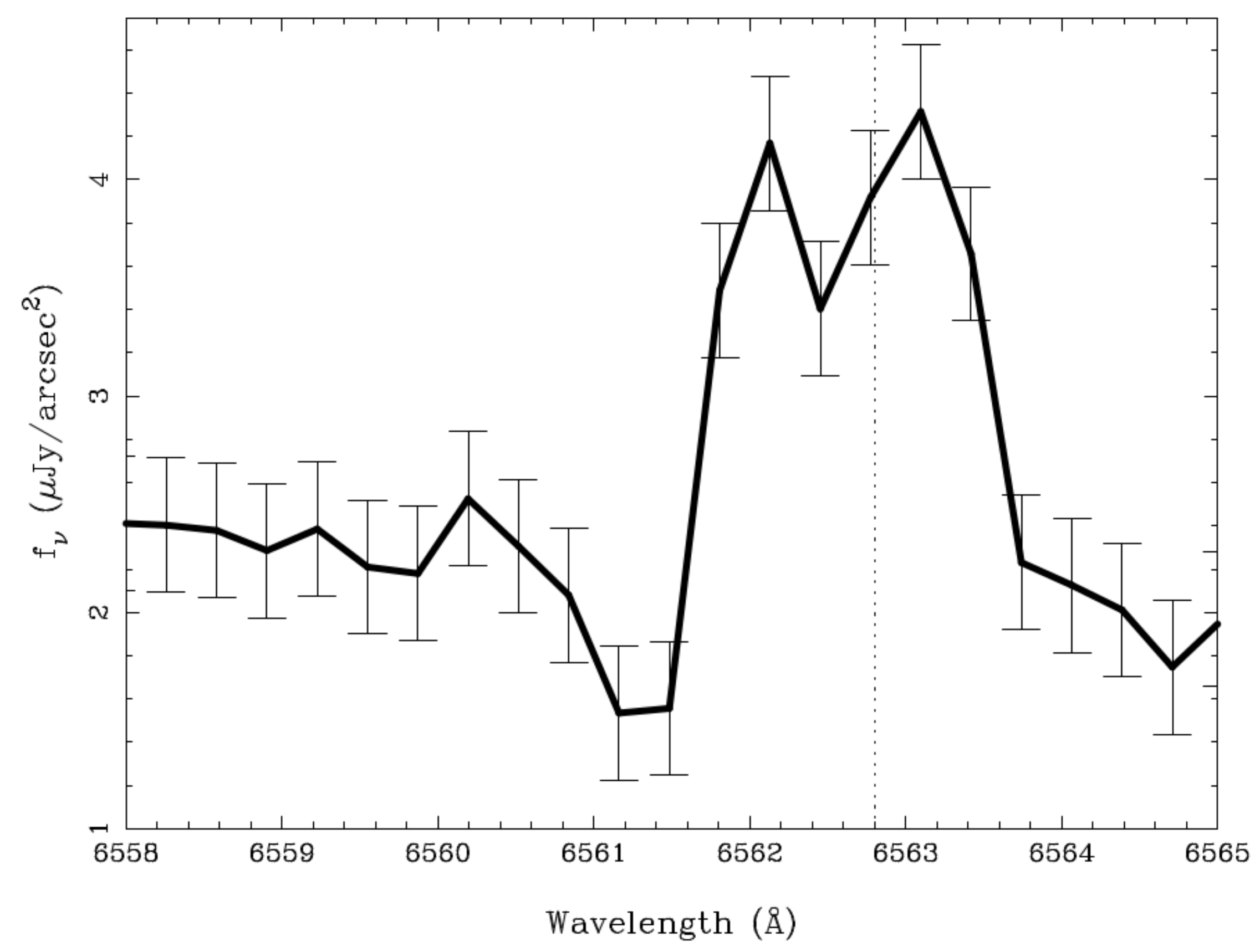}
\caption{H$\alpha$ spectrum of the sky from the South-East side of the
  V1315 Aql slit. The error bars show the noise levels of the
  background sky}
\label{fig:hazoom}
\end{figure}

There is also H$\alpha$ emission present in both the sky portions of
the slit centred on V1315 Aql, though any N[II] lines present are lost
in the noise. We show the H$\alpha$ line profile from the sky on the
South-East side of the V1315 Aql slit in Figure \ref{fig:hazoom}.

The FWHM of the H$\alpha$ and N[II] lines of all the shell slits are
listed in Table \ref{tab:fwhm}.

\begin{table*}
  \caption[]{The first column shows the radial velocities in km
    s$^{-1}$ of the H$\alpha$ emission line in the spectra of the
    seven V1315 Aql shell slits and the sky portion of the four
    flux calibration stars. The errors on the velocities are $\pm5$
    km\,s$^{-1}$. The other columns shows the FWHM in km s$^{-1}$ of
    each line. The errors on the FWHM are $\pm8$ km\,s$^{-1}$. The
    H$\alpha$ line in the V1315 Aql sky has a double peak and both
    radial velocities are shown. All values and errors were obtained
    from Gaussian fits to the emission lines.}
\begin{center}
\begin{tabular}{lrrrrrrrrrr}
\hline
\multicolumn{1}{l}{Slit name} & \multicolumn{1}{r}{Radial} & \multicolumn{1}{r}{H$\alpha$} & \multicolumn{1}{r}{N[II]} & \multicolumn{1}{r}{N[II]} & \multicolumn{1}{r}{H$\beta$} & \multicolumn{1}{r}{NI} & \multicolumn{1}{r}{NII} & \multicolumn{1}{r}{NI} & \multicolumn{1}{r}{SII} & \multicolumn{1}{r}{SII} \\
&  \multicolumn{1}{r}{Velocity} & & \multicolumn{1}{r}{6548\AA} & \multicolumn{1}{r}{6583\AA} & & \multicolumn{1}{r}{5676\AA} & \multicolumn{1}{r}{5740\AA} & \multicolumn{1}{r}{5742\AA} & \multicolumn{1}{r}{6716\AA} & \multicolumn{1}{r}{6731\AA} \\
\hline
Shell 1 & $-8$ & 71 & 82 & 72  & 0 & 0 & 0 & 0 & 0 & 0 \\
Shell 2 & $-8$ & 86 & 65 & 74  & 0 & 0 & 0 & 0 & 67 & 78 \\
Shell 3 & $-5$ & 80 & 79 & 79  & 122 & 143 & 206 & 0 & 68 & 68 \\
Shell 4 & $-3$ & 79 & 59 & 75  & 89 & 78 & 0 & 124 & 0 & 0 \\
Shell 5 & $-9$ & 75 & 54 & 75 & 0 & 0 & 0 & 0 & 52 & 64 \\
Shell 6 & $-26$ & 84 & 87 & 80 & 0 & 0 & 0 & 0 & 78 & 76 \\
Shell 7 & 4 & 89 & 87 & 86 & 0 & 0 & 0 & 0 & 91 & 84 \\
V1315 Aql Sky& $-33$ \& 14 & 0 & 0 & 0 & 0 & 0 & 0 & 0 & 0 & 0 \\
Star 1 & 11 & & & & & & & & & \\
Star 2 & $-23$ & & & & & & & & & \\
Star 3 & 29 & & & & & & & & & \\
Star 4 & 14 & & & & & & & & & \\
\hline
\end{tabular}
\end{center}
\label{tab:fwhm}
\end{table*}

\subsubsection{Systemic velocity of V1315 Aql}

Historically the systemic velocity of V1315 Aql, $\gamma$, has been
difficult to determine because of the complex behaviour of its disk
emission lines and lack of absorption lines from the primary and
secondary stars. \citet{downes86} presented radial velocity data for
H$\beta$, H$\gamma$ and HeII 4686\AA\, emission lines. They
derived values for $\gamma$ consistent with zero from the H$\beta$ and
HeII 4686\AA\, lines, but the H$\gamma$ line gave a value of 100
km\,s$^{-1}$. \citet{dhillon91} also used the H$\beta$, H$\gamma$ and
HeII 4686\AA\, emission lines and the HeI 4471\AA\, line and
derived a $\gamma$ range of $-4$ to +93 km\,s$^{-1}$. Given the
unreliability of the broad emission lines from the accretion disc to
determine $\gamma$, in the following subsection we will use our own
measurements of the radial velocity of the shell to determine if they
are consistent.

\subsubsection{Radial velocities of shell emission lines}

To measure the radial velocities of the emission lines, we fitted a
Gaussian to the H$\alpha$ line of the shell and measured the
wavelength at the centre of the Gaussian. The resulting shell radial
velocities are shown in Table \ref{tab:fwhm}. 

The spectrum of the sky on the South-East side of the slit centred on
V1315 Aql is shown in Figure \ref{fig:hazoom}. The plot shows
tentative evidence of a double-peaked structure. We measured the
radial velocity of each peak to be $-33$ and 14 km\,s$^{-1}$. If we
assume that the two peaks represent emission from the front and back
sides of a spherically-expanding shell, then the average of the two
gives a systemic velocity of $ \gamma \approx -10$ km\,$^{-1}$, and an
expansion velocity of $\sim$ 25 km\,s$^{-1}$. We analysed the sky on
the North-East side of V1315 Aql and it too showed a double-peaked
structure, although it is less pronounced.

The seven shell slits were placed at the edges of the shell. The
expansion velocity of the edge of the shell will be tangential to the
line of sight and will not affect the radial velocities, which should
be similar to the overall systemic velocity. The measured shell radial
velocities are shown in Table \ref{tab:fwhm} and are broadly
comparable with the systemic velocity of $-10$ km\,s$^{-1}$ derived
above, apart from shells 6 and 7 which differ by 14 and 16 km\,s$^{-1}$
respectively.

The Galactic velocity of V1315 Aql relative to the Sun can be
derived from its Galactic coordinates, $l=46.4^{\circ}, b=0^{\circ}$,
which give a radial velocity of 7 km\,s$^{-1}$. This is broadly
consistent with the systemic velocity derived above.

\subsubsection{Line fluxes}

The fluxes of the emission lines from the shell in each of the slits
are given in Table \ref{tab:flux}. Assuming a shell radius of
220$\arcsec$, a distance of 489 parsecs \citep{ak2008}, and using the
H$\alpha$ flux from each slit, we can estimate the total flux from the
whole shell. However, we can see by examining Figure \ref{fig:geocirc}
that the shell is fragmented and clumpy and only a small fraction is
actually emitting. If we assume that 10\% of the full shell is
emitting and take an average H$\alpha$ flux from the seven shell slits
of $1.70 \times 10^{-17}$ ergs/cm$^2$/sec/arcsec$^{2}$, we obtain a total
H$\alpha$ luminosity of $7.1 \times 10^{30}$ ergs/sec.

The plots of \citet{downes01} showing the temporal reduction in the
H$\alpha$ luminosities of shells from fast and slow novae, have a
lower limit of log L = 30 at 100 yrs. We note that the source of the
V1315 Aql luminosity is likely to include emission from shock
interaction with pre-existing ISM. This would enhance the flux, and
lead to an underestimate of the age of the shell. We conclude that the
shell is likely to be significantly older than 100 yrs.

\begin{table}
\caption[]{H$\alpha$ and N[II] flux (ergs/s/cm$^2$/arcsec$^2$ $\times$
  $ 10^{-18}$) from the seven shell slits. The errors on the flux
  values are $\pm 25\%$. }
\begin{center}
\begin{tabular}{crrr}
  \hline
  \multicolumn{1}{l}{Shell slit No.} &
  \multicolumn{1}{r}{H$\alpha$} & \multicolumn{1}{r}{N[II]} &
  \multicolumn{1}{r}{N[II]} \\ \multicolumn{1}{l}{} &
  \multicolumn{1}{l}{} & \multicolumn{1}{r}{6548\AA} &
  \multicolumn{1}{r}{6583\AA} \\ \hline 1 & 17.2 & 4.22 & 10.8 \\ 2 &
  18.1 & 2.10 & 7.64 \\ 3 & 18.3 & 4.16 & 11.8 \\ 4 & 24.2 & 1.47 &
3.03 \\ 5 & 9.58 & 0.73 & 3.62 \\ 6 & 15.4 & 2.64 & 5.61 \\ 7 & 16.2
& 4.80 & 11.9 \\
\hline
\end{tabular}
\end{center}
\label{tab:flux}
\end{table}

\subsubsection{Time of nova eruption}
\label{time}

The distance to V1315 Aql was measured by \citet{ak2008} as $489 \pm
49$ pc, computed from the Period-Luminosity-Colours (PLCs) relation of
CVs calibrated with {\em 2MASS} photometric data. The angular radius
of the shell on our image is $\sim$4$\arcmin$, giving a physical radius
of $1.7 \times 10^{13}$ km. \citet{duerbeck87} found that the velocity of
nova shells reduces by half every 50--100 yrs.  Using our measured
expansion velocity of $\sim$25 km\,s$^{-1}$, and assuming an initial
velocity of 2,000 km\,s$^{-1}$ (see Table 8.1 \citealt{bode08}), we
estimate that the nova explosion occurred $\sim$500-600 yrs
ago. However, if we take more extreme values for the initial ejection
velocity, say 700 km\,s$^{-1} $ and a deceleration half-life of 200
yrs then the age of the nova increases to 1,200 yrs.

Assuming the visual magnitude of V1315 Aql was the same prior to the
nova event as it is now, m$_{V}=14.3$, and taking the average
brightening of a nova to be $\sim$11 magnitudes (\citealt{bode08}),
the system would have been at m$_{V}\sim$ 3.3 at peak brightness,
clearly visible to the naked eye. Novae decline rapidly and so it
would have dropped below the naked-eye visibility limit of m$_{V}\sim$
6 within a few days. We reviewed the catalogues of ancient Chinese and
Asian novae and supernovae sightings by \citet{stephenson76} which
includes sightings from 532 BC up to 1604 AD. We could find no record
of an event close to the coordinates of V1315 Aql. If the nova
eruption occurred when V1315 Aql was close to the Sun in the sky, and
it was brighter than m$_{V}\sim$ 6 for only a few days, it may well
have been hidden in twilight and hence gone unnoticed.

\subsubsection{Temperature and density of the shell}

The method most often used to determine the temperature and density of
gaseous nebulae is to measure the ratio of the intensities of
particular emission lines from the same species of ions. Two ions
which are often used are N[II] and O[III]. We were unable
to detect any O[III] lines in our spectra, and the N[II] ratio
requires a flux measurement of the 5755\AA\,\,line, which we were only
able to detect very weakly in shell slit 1. It was not present in any
other slit. Hence we can only place an upper limit on the electron
temperature ($T_e$) of the shell of 5,000\,K using Figure 5.1 from
\citet{osterbrock89}.

\subsubsection{Mass of the shell}

We can derive a rough estimate of the mass of the shell using the
technique set out in \citet{corradi2015}. They derived the ionised
hydrogen masses of several planetary nebulae using the formula
\begin{equation}
  m_{\mathrm{shell}}(H^{+})=\frac{4 \pi
    \,D^2\,F(\mathrm{H}\beta)\,m_{\mathrm{p}}}{h\nu_{\mathrm{H}\beta}\,n_\mathrm{e}\,\alpha_{\mathrm{H}\beta}^{eff}(H^0,T_\mathrm{e})},
\end{equation}
where \textit{D} is the distance to the object, \textit{F}(H$\beta$)
is the H$\beta$ flux, \textit{m}$_{\mathrm{p}}$ is the mass of a
proton, \textit{h$\nu_{\mathrm{H}\beta}$} is the energy of an H$\beta$
photon, \textit{n$_{\mathrm{e}}$} is the electron density per cm$^3$,
and \textit{$\alpha_{\mathrm{H}\beta}^{eff}(H^0,T_{\mathrm{e}})$} is
the effective recombination coefficient for H$\beta$. This formula is
also applicable to nova shells \citep{osterbrock89}.

As we pointed out in Section \ref{emline}, the spectra of our four
flux calibration stars do not cover H$\beta$ so we are unable to
derive a flux directly. However, we can make a rough estimate as
follows. The H$\beta$ line is present in four shell slits (Nos. 3, 4,
6 and 7). We can measure the counts for both H$\alpha$ and
H$\beta$. The DEIMOS exposure time calculator for a source that is
flat in frequency gives the ratio of counts for H$\alpha$\,:\,H$\beta$
as approximately 1\,:\,0.3. Assuming that 10\%\ of the full shell is
emitting and taking an average H$\alpha$ flux from the seven shell
slits of $1.70 \times 10^{-17}$ ergs/cm$^2$/sec/arcsec$^{2}$ we obtain
a total H$\alpha$ flux of $2.49 \times 10^{-13}$ ergs/cm$^2$/sec from the
whole shell, allowing us to derive an H$\beta$ flux of
\textit{F(H$\beta$)} =$ 8.9 \times 10^{-14}$ ergs/cm$^2$/sec.

The electron density, \textit{$n_e$}, can be estimated using the S[II]
6716 and 6731 line ratio, which we found to be 1.4 (see Section
\ref{emline}). Figure 5.8 in \citet{osterbrock89} shows the electron
density versus intensity ratio at \textit{T$_{\mathrm{e}}$} =
10,000\,K and indicates a scaling of
\textit{n$_e$}(10$^4/$\textit{T$_{\mathrm{e}}$})$^{1/2} $. We found a
maximum temperature of 5,000\,K which gives an electron density of
$\sim$ 22 cm$^{-3}$.

Finally, using the distance measured by \citet{ak2008} of 489 pc and a
value for $\alpha_{\mathrm{H}\beta}^{eff}(H^0,T_\mathrm{e})$ of $3.78
\times 10^{-14}$, for Case A conditions at T$_{\mathrm{e}}$ = 5000\,K
listed in Table 4.1 of \citet{osterbrock89} gives a maximum mass of
\begin{equation}
  m_{\mathrm{shell}}(H^{+}) \simeq 2 \times 10^{-4} M_{\odot}.
\end{equation}

There is no need to correct for extinction as \citet{rutten92b} found
$E(B-V)=0$ for V1315 Aql using \textit{IUE} spectra of interstellar
absorption bands around 2200\AA. In view of the many assumptions used
to estimate the mass of the shell, it should be treated as an order of
magnitude approximation.

As nova shells expand they decelerate as they sweep up pre-existing
circumstellar gas, which leads to a doubling of their mass every
50--100 yrs \citep{duerbeck87}. We estimated the age of the shell in
Section \ref{time} as 500--1200 yrs, so the original ejected mass of
the shell would have been substantially lower than the value we have
derived above, giving a maximum ejected mass of $\la 10^{-5}$
M$_{\odot}$. This rules out a planetary nebula origin, which typically
have masses in the range 0.1--1.0 M$_{\odot}$
\citep{osterbrock89}. Nova shells typically have masses in the range
10$^{-4}$--10$^{-6}$~M$_{\odot}$ (\citealt{yaron05}), so our estimate
of the shell mass in V1315 Aql of $\sim 10^{-5}$ M$_{\odot}$ is in
accordance with this.

\section{Discussion}
\label{sec:disc}

We can summarise our findings as follows. The shell is broadly
spherical and appears to be centred on V1315 Aql, strongly suggesting
that the shell is associated with the central binary. The systemic
velocity of the shell measured from the sky portion of the V1315 Aql
slit and at the edges of the shell are broadly consistent. The absence
of 22$\mu$m emission precludes a planetary nebula origin
(\citealt{mizuno10}). We derive an order-of-magnitude estimate of the
mass of the shell of $\sim 10^{-5}$ M$_{\odot}$ which rules out a
planetary nebula or supernova origin. We conclude that these results
indicate that the shell is associated with V1315 Aql.

At this stage of the shell's evolution, the luminosity of the outer
edges of the shell is most likely fuelled by two processes,
recombination and shock interaction with pre-existing CSM. Our flux
measurement will include contributions from both of these processes,
making it difficult to estimate the physical conditions in the shell
as a whole. Furthermore, the lack of other forbidden emission lines in
the shell spectra, especially N[II] 6583\AA\,\,and O[II], means we
cannot determine the physical parameters of the shell to confirm
conclusively that it exhibits properties consistent with a nova
origin.

In S15, we estimated that the nova-like phase following a nova
eruption lasts $\sim2400$ yrs. This is comparable to the $\sim2000$
yrs order-of-magnitude estimate by \citet{patterson13}, based on the
transition of BK Lyn to a dwarf nova in the year 2011. However, the
AAVSO light curve of BK Lyn\footnote{https://www.aavso.org/} suggests
that it has now reverted back to a nova-like state, indicating that
the object is a Z Cam-type dwarf nova that has likely been
transitioning from the nova-like to dwarf nova state for much less
than the $\sim2000$-yr estimate of \citet{patterson13}.
\citet{shara17a} found that the transition time for AT Cnc was much
shorter at $330^{+135}_{-90}$ yrs. Our estimate of the time since
the nova eruption on V1315 Aql of 500--1200 yrs is consistent with
both these timescales, and lies within the overall nova recurrence
timescale of 13000 yrs found by \citet{schmidtobreick15}.
 
\section{Conclusions}

We present images and spectra of the shell surrounding V1315 Aql. Our
results strongly suggest that the shell originated from a nova
eruption on the CV. This discovery of the first nova shell around a
nova-like variable adds further support to the theory of nova-induced
cycles in the mass transfer rates of CVs.


\section*{Acknowledgments}

We would like to thank the referee for his helpful comments and for
pointing out the latest AAVSO light curve of BK Lyn demonstrating Z
Cam-like behaviour.

VSD and SPL were supported under grants from the Science
and Technology Facilities Council (STFC).
This publication makes use of VOSA, developed under the Spanish
Virtual Observatory project supported from the Spanish MICINN through
grant AyA2011-24052. The INT is operated on the island of La Palma by
the Isaac Newton Group of Telescopes in the Spanish Observatorio del
Roque de los Muchachos of the Instituto de Astrof\'{\i}sica de
Canarias. Some of the data presented herein were obtained at the
W.M. Keck Observatory, which is operated as a scientific partnership
among the California Institute of Technology, the University of
California and the National Aeronautics and Space Administration. The
Observatory was made possible by the generous financial support of the
W.M. Keck Foundation. The authors wish to recognise and acknowledge
the very significant cultural role and reverence that the summit of
Mauna Kea has always had within the indigenous Hawaiian community.  We
are most fortunate to have the opportunity to conduct observations
from this mountain.


\bibliographystyle{mn2e} \bibliography{abbrev.bib,refs.bib}

\label{lastpage}

\end{document}